\documentclass[prb,nofootinbib,twocolumn,superscriptaddress]{revtex4} 

\usepackage{graphicx}
\usepackage{dcolumn}
\usepackage{bm}
\usepackage{threeparttable}
\usepackage{times}
\usepackage{mathptmx}
\usepackage{lscape}
\usepackage{natbib}
\usepackage{amsmath}
\usepackage{amssymb}
\usepackage{braket}
\usepackage{comment}
\usepackage{color}

\def\degree{\kern-.2em\r{}\kern-.3em}

\begin{document}


\title{  Nonlinearity in Canonical Ensemble for Multicomponent Alloys\\ Revisited from Structural Degree of Freedoms  }

\author{Koretaka Yuge}
\affiliation{
Department of Materials Science and Engineering,  Kyoto University, Sakyo, Kyoto 606-8501, Japan\\
}%

\begin{abstract}
{  For classical discrete system under constant composition typically referred to substitutional alloys, we examine local nonlinearity in canonical average $\phi$.
We have respectively investigated the local and global behavior of nonlinearity through previously-introduced vector field $\mathbf{A}$ and through tropical limit of the vector field. While these studies indicated the importance of constraints to structural degree of freedoms (SDFs) for global nonlinearity, it has been still unclear how the constraints to SDF affects local nonlinearity.
Based on statistical manifold, we make intuitive bridge between the SDF-based information and local nonlinearity, decomposing the local nonlinearity into two (for binary alloys with pair correlations) or three (for otherwise) contributions in terms of the Kullback-Leibler divergence, where this decomposition is independent of temperature and many-body interaction, and is defined on individual configuration. 
We also find that we can provide $\mathbf{A}$-dependent as well as $\mathbf{A}$-independent decomposition of the local nonlinearity, where non-separability in SDFs and its nonadditive character is independent of $\mathbf{A}$, which indicates that information about \textit{evolution} of the vector field should be required to address the non-separability and nonadditivity. 
The present work enables to quantify how configuration-dependent constraints to SDF affect local nonlinearity in canonical average for multicomponent alloys.  }
\end{abstract}


\maketitle

\section{Introduction}
For classical discrete system under constant composition with $f$ structural degree of freedoms (SDFs), typically referred to substitutional alloys, expectation value of structure along chosen coordination $p$ under given coordination $\left\{ Q_{1}, \cdots, Q_{f} \right\}$ is provided through canonical average, namely,
\begin{eqnarray}
\label{eq:can}
\Braket{ Q_{p}}_{Z} = Z^{-1} \sum_{i} Q_{p}^{\left( i \right)} \exp \left( -\beta U^{\left( i \right)} \right),
\end{eqnarray}
where $\Braket{\quad}_{Z}$ denotes canonical average, $U$ potential energy, $Z=\sum_{i}\exp\left(-\beta U^{\left( i \right)}\right)$ partition function, $\beta$ inverse temperature, and summation $i$ is taken over all possible microscopic states on configuration space (i.e., configurations). 
One of the most natural selection for the coordination is employing complete orthonormal basis (COB), which is typically constructed by generalized Ising model (GIM),\cite{ce} considered in the present case. When we employ such COB, potential energy $U$ for any configuration $k$ is exactly given by
\begin{eqnarray}
\label{eq:u}
U^{\left( k \right)} = \sum_{j=1}^{f} \Braket{U|Q_{j}} Q_{j}^{\left( k \right)},
\end{eqnarray}
where $\Braket{\quad|\quad}$ denotes inner product, i.e., trace over possible states on configuration space: 
$\Braket{f|g} = \rho^{-1} \sum_{i} f^{\left( i \right)} \cdot g^{\left( i \right)}$, with
$\rho$ representing normalization constant for the inner product. 
When we introduce two $f$-dimensional vectors of $\mathbf{Q}_{Z}=\left(\Braket{ Q_{1}}_{Z},\cdots, \Braket{ Q_{f}}_{Z}\right)$ and $\mathbf{U}=\left(U_{1},\cdots,U_{f}\right)$ ($U_{b}=\Braket{U|Q_{b}}$), 
canonical average of Eq.~\eqref{eq:can} can be interpreted as the following map $\phi$:
\begin{eqnarray}
\label{eq:maps}
\phi: \mathbf{U} \mapsto \mathbf{Q}_{Z} = \phi\cdot\mathbf{U}.
\end{eqnarray}
It is clear from Eqs.~\eqref{eq:can}-\eqref{eq:maps} that generally, $\mathbf{Q}_{Z}$ is not a linear function of $\mathbf{U}$, i.e., canonical average $\phi$ is a nonlinear map. 
Due to the complicated nonlinear character, it is typically difficult to exactly determine temperature dependence of $\Braket{Q}_{Z}$ for multicomponent alloys: Thus, various approaches have been developed including Metropolis algorism, entropic sampling and Wang-Landau sampling for efficient exploration of important microscopic configurations to determine equilibrium properties.\cite{mc1,mc2,wl} 

Very recently, we quantitatively formulate local nonlinearity of the canonical average $\phi$, through the newly-introduced vector field $\mathbf{A}$ that is essentially independent of temperature and energy.\cite{bd} 
By applying tropical limit to the vector field, we also investigated how global nonlinearity is governed in terms of lattice geometry, finding that linear region for $\phi$ around disordered state is roughly bounded by spatial constraint to individual SDF.\cite{trop} 
Although these facts indicate that constraint conditions to SDFs should play essential role on the nonlinearity in $\phi$, their link is still unclear. 
We here show that statistical manifold enables geometric intuition of how local nonlinearity in $\phi$ is interpreted based on the introduced vector field $\mathbf{A}$ and information about SDFs, and how $\mathbf{A}$ and SDFs are linked. The details are shown below. 

\section{Derivation and Concepts}
Let us first briefly see local nonlinearity in $\phi$ on configuration space, which has been partly addressed in our recent studies. 
Hereinafter for simplicity (without loss of generality), structure (value of $Q$s) is measured from that at origin (i.e., center of gravity) of configurational density of states (CDOS), given by $\left\{\Braket{Q_{1}},\cdots,\Braket{Q_{f}}\right\}$, where $\Braket{\quad}$ denotes taking linear average for all possible microscopic configurations: $\Braket{f}=N^{-1}\sum_{i=1}^{N}f^{\left( i \right)}$. Note that hereinafter, CDOS always means density of configurations in terms of GIM coordinates $\left\{ Q_{1},\cdots, Q_{f} \right\}$ \textit{before} applying many-body interaction to the system.
If the canonical average $\phi$ is a globally linear map, it can be given by the following form of 
\begin{eqnarray}
\label{eq:lin}
\phi = -\beta \cdot \Lambda , 
\end{eqnarray}
where $\Lambda$ is temperature-independent $f\times f$ matrix, and $\beta$ is required from dimensional analysis between $\mathbf{U}$ and $\mathbf{Q}_{Z}$. 
Our previous study finds that when CDOS takes multidimensional gaussian, $\phi$ always becomes a linear map for any given many-body interaction,\cite{em2} which provides $\Lambda$ as invertible map.
On the other hand, when we perform series expansion of canonical average in terms of moments, we get
\begin{widetext}
\begin{eqnarray}
\label{eq:spn}
\mathrm{M}\left[\Braket{Q_{p}}_{Z} \right]= \sum_{n=1}^{\infty} \left[  \frac{\left( -1\right)^{n}}{ n!}  \sum_{k_{1},\cdots,k_{n}} u_{k_{1}}\cdots u_{k_{n}}  \left( \Braket{Q_{p} Q_{k_{1}}\cdots Q_{k_{n}}} - \sum_{j_{1},\cdots, j_{n+1}}\Braket{Q_{j_{1}}\cdots Q_{j_{k}}}\cdots \Braket{Q_{j_{m}}\cdots Q_{j_{n+1}}}  \right)   \right],
\end{eqnarray}
\end{widetext}
where $u_{k} = \beta \Braket{U|Q_{k}}$ and $\Braket{Q_{k1}\cdots Q_{kn}}$ corresponds to the $n$-th order moment for CDOS. From Eqs.~\eqref{eq:lin} and \eqref{eq:spn}, it is clear that in order to make $\phi$ as a linear map in terms of the series expansion, relationships in CDOS between $t$-th ($t\ge 3$) order moments and second-order moments of $\left\{ \Braket{Q_{i}Q_{k}}|i,k=1,\cdots,f \right\}$ should be same as the relationships for multidimensional Gaussian. 
Therefore, we here naturally introduce measure for nonlinearity in $\phi$ for discrete systems in terms of any deviation in practical CDOS from multidimensional gaussian that has the same covariance matrix $\Gamma$ as practical CDOS.  This reminds us important fact that nonlinearity in $\phi$ is  a priori determined only by information about CDOS landscape on given lattice. 

We recently introduce another insight into local nonlinearity in $\phi$, called ``anharmonicity in structural degree of freedom (ASDF)'' , which is a vector field defined on individual configuration $\mathbf{Q}$:\cite{asdf}
\begin{eqnarray}
\label{eq:asdf}
\mathbf{A}\left( \mathbf{Q} \right) &=& \left( A_{1}\left( \mathbf{Q} \right),\cdots,A_{f}\left( \mathbf{Q} \right) \right) \nonumber \\
A_{i}\left( \mathbf{Q} \right) &=&  \left\{\left(\phi\left( \beta \right)\circ \left( -\beta\cdot\Lambda \right)^{-1} \right)\cdot \mathbf{Q} - \mathbf{Q}\right\}_{i}.
\end{eqnarray}
Here, $\left\{ \quad \right\}_{i}$ represents $i$-th component of $\mathbb{R}^{f}$ vector, and  
when $\phi$ acts as locally linear map, i.e., $\phi=-\beta\cdot\Lambda$ around $\mathbf{Q}$, $\mathbf{A}$ clearly takes zero vector.
Since we have shown that $\Lambda$ is an invertible linear map and image of composite map $\phi\left( \beta \right)\circ \left( -\beta\cdot\Lambda \right)^{-1} $ is exactly independent of energy and of temperature,\cite{asdf, bd} $\mathbf{A}$ is a measure of nonlinearity in $\phi$ depending only on CDOS landscape, and independent of temperature and many-body interaction, which is a desired property as 
discussed above: i.e., ASDF can be known a priori without any thermodynamic information. 
The importance of the ASDF $\mathbf{A}$ has been demonstrated, which can quantitatively formulate bidirectional stability relationships $B$ in canonical average, i.e., $f$-dimensional hypervolume ratio between a set of $\mathbf{Q}_{Z}$ and a set of $\mathbf{U}$:\cite{bd}
$B= \log \left| 1+ \mathrm{div} \mathbf{A} + \sum_{\mathrm{F}}J_{\mathrm{F}}\left[ {\partial \mathbf{A}}/{ \partial Q} \right]  \right|$,
where $J$ is Jacobian, and summation is taken over all subspaces on configuration space considered. 

In order to further investigate the nonlinearity extending from local to global behavior, we recently apply tropical limit to ASDF with $f=1$ pair correlation system, i.e., $\lim_{t\to\infty}\log_{t}\mathbf{A}$.  
The result shows that tropical limit of linear region for $\phi$ on configuration space exists around origin (i.e., perfectly disordered state):
\cite{trop} 
\begin{eqnarray}
\label{eq:tropical}
 M\gamma < \Omega_{a} < -c\cdot M\gamma,
\end{eqnarray}
where $\Omega_{a}$ corresponds to tropical limit for coordination $Q_{a}$, and $\gamma$ and $c\left( \left|c\right|<1 \right)$ respectively takes negative and positive constant, and $M$ corresponds to exponent of number of simple cycles consisting of considered pairs. 
Eq.~\eqref{eq:tropical} indicates that linear character in $\phi$ can be enhanced by decreasing spatial constraint on constituents (corresponding to increase value of $M$), i.e., by decreasing constraints on individual structural degree of freedom (SDF) dominated by underlying lattice. 

\begin{figure}[h]
\begin{center}
\includegraphics[width=1.05\linewidth]{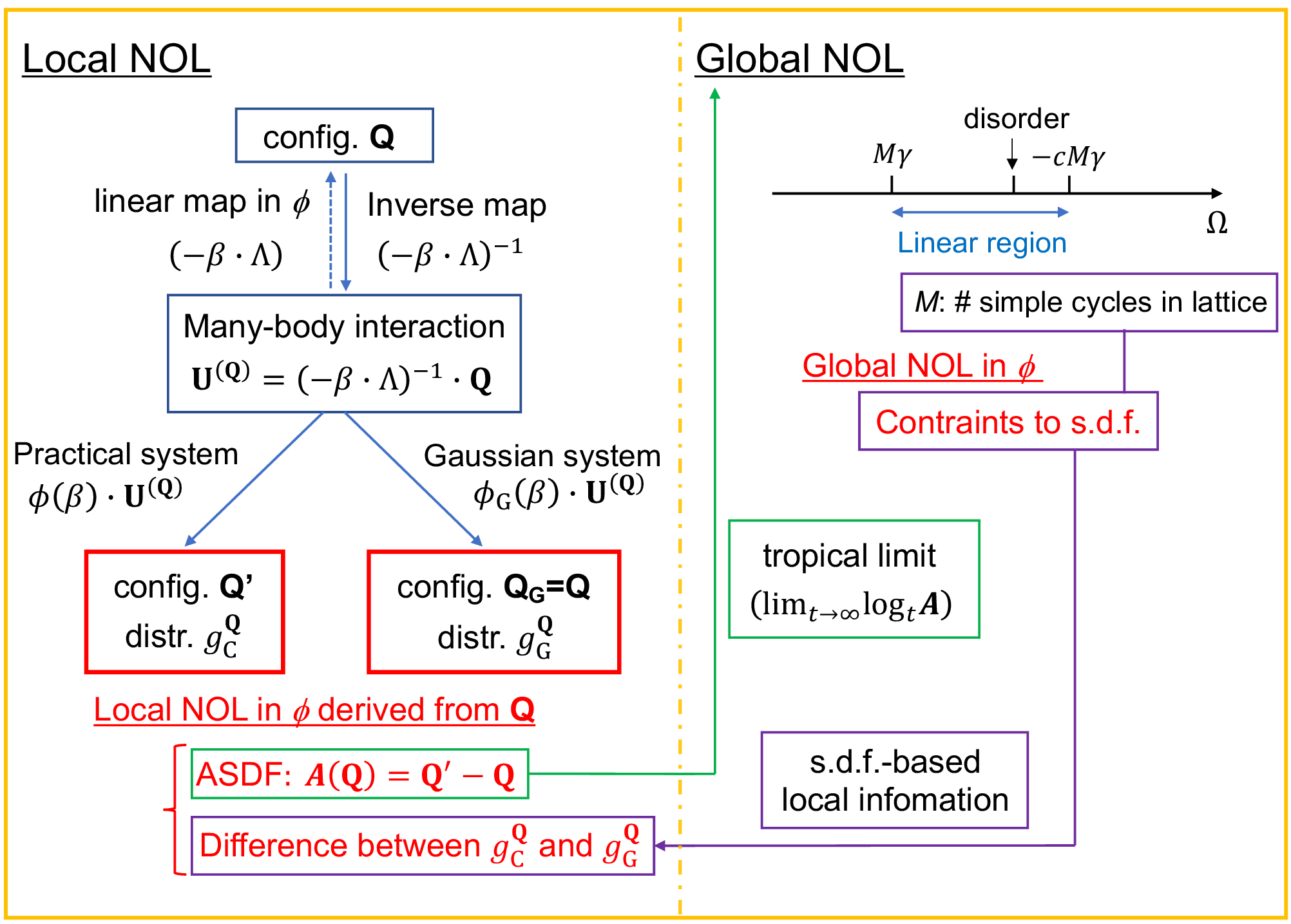}
\caption{ Concept for nonlinearity in canonical ensemble.}
\label{fig:concept}
\end{center}
\end{figure}

From above discussions, we see that (i) the local nonlinearity in $\phi$ can be quantitatively investigated on individual configuration by ASDF of $\mathbf{A}$, which is independent of temperature and energy, (ii) nonlinearity can come from deviation in CDOS from multidimensional gaussian with the same covariance matrix of CDOS, and (iii) global behavior for the nonlinearity decreases by decreasing constraint to SDF. 
The concepts of global and local nonlinearity in $\phi$ for above discussions are summarized in Fig.~\ref{fig:concept}, where we define 
canonical average for Gaussian CDOS (called Gaussian system) as $\phi_{\textrm{G}}$, and also define canonical distribution under potential energy $\mathbf{U^{\left( Q \right)}} = \left( -\beta\Lambda \right)^{-1}\cdot \mathbf{Q}$ for practical and Gaussian system as $g_{\textrm{C}}^{\mathbf{Q}}$ and $g_{\textrm{G}}^{\mathbf{Q}}$. Note that (i) both distributions of $g_{\textrm{C}}^{\mathbf{Q}}$ and $g_{\textrm{G}}^{\mathbf{Q}}$ are independent of temperature and energy, from the character of ASDF, and (ii) center of gravity for $g_{\textrm{C}}^{\mathbf{Q}}$ and $g_{\textrm{G}}^{\mathbf{Q}}$ respectively corresponds to $\mathbf{Q}'$ and $\mathbf{Q}_{\textrm{G}}\left( =\mathbf{Q} \right)$ with $\mathbf{A}\left( \mathbf{Q} \right) = \mathbf{Q}' - \mathbf{Q}_{\textrm{G}}$.
From Fig.~\ref{fig:concept}, main purpose of the present study is to make intuitive bridge between 
constraints for SDFs, bounding global behavior of nonlinearity in $\phi$ (r.h.s. of Fig.~\ref{fig:concept}) and local nonlinearity derived from configuration $\mathbf{Q}$ (l.h.s. of Fig.~\ref{fig:concept}), by introducing proper measure for local nonlinearity, under the desired condition of temperature- and energy-independent formulation.
It appears difficult to extract information about constraints to SDFs from a single ASDF itself, since it corresponds to a difference between \textit{averaged} information for canonical distributions of $g_{\textrm{C}}^{\mathbf{Q}}$ and $g_{\textrm{G}}^{\mathbf{Q}}$. 
We therefore examine the local nonlinearity derived from configuration $\mathbf{Q}$ in terms of SDFs, by employing statistical manifold\cite{ig} to measure the difference among multiple discretized canonical distributions. 

Here, we start from discrete binary system on lattice with $f=2$ under \textit{constant} composition, consisting of two pair correlations, which is a minimal but an essential system to investigate relationships between nonlinearity and constraints to SDFs. 
We express any canonical distribution as 
\begin{eqnarray}
g\left( x,y \right) &=& \sum_{i=0}^{n}\sum_{k=0}^{m}g_{ik}X_{i}\left( x \right)Y_{k}\left( y \right), \nonumber \\
X_{i}\left( x \right)&=&\delta_{x{Q}_{1i}} \quad Y_{k}\left( y \right)=\delta_{y{Q}_{2k}},
\end{eqnarray}
 with $Q_{jm}$ representing $j$-th pair correlation at $m$-th discretized grid. In this case, covariance matrix of CDOS for practical system becomes diagonal at thermodynamic limit, which naturally results in that canonical distribution obtained from corresponding Gaussian system is separable, i.e., $g_{\textrm{G}}^{\mathbf{Q}}\left( x,y \right)=g_{\textrm{G}}^{\mathbf{Q}}\left( x \right)\cdot g_{\textrm{G}}^{\mathbf{Q}}\left( y \right)$. 
Then, any canonical distribution can always be explicitly rewritten as a member of exponential family,\cite{ig} namely
\begin{eqnarray}
g\left( x,y \right) &=& \exp\left\{  \left( \sum_{\left( i,k \right) \neq \left( 0,0 \right)}^{\left( n,m \right)}  \tilde{\theta}_{ik} X_{i} Y_{k}    \right) - \Psi\left( \tilde{\theta} \right)  \right\}, \nonumber \\
\tilde{\theta}_{ik} &=& \log \frac{g_{ik}}{g_{00}}, \nonumber \\
\Psi\left( \tilde{\theta} \right) &=& -\log g_{00} = \log \left\{ 1+ \sum_{\left( i,k \right) \neq \left( 0,0 \right)}^{\left( n,m \right)} \exp \left( \tilde{\theta}_{ik} \right) \right\}. 
\end{eqnarray}
Under these definitions, $\left\{ \tilde{\theta} \right\}$ corresponds to affine coordinate and its dual affine coordinate is provided through Legendre transform of 
$\tilde{\eta}_{ik} = {\partial \Psi\left( \tilde{\theta} \right)}/{ \partial \tilde{\theta}_{ik}} = g_{ik} = g_{00}\cdot\exp \tilde{\theta}_{ik}$.
In order to explicitly treat nonlinearity in terms of constraints to SDF, we here employ another coordinate by performing affine transform to retain $e$- and $m$-flatness:
\begin{eqnarray}
\label{eq:af}
\theta_{i} &=& \tilde{\theta}_{i0} \quad \left( 1\le i \le n \right)\nonumber \\
\theta'_{k} &=& \tilde{\theta}_{0k} \quad \left( 1\le k \le m \right)\nonumber \\
\theta_{ik} &=& \log \frac{g_{00} \cdot g_{ik}}{ g_{i0} \cdot g_{0k} } = \tilde{\theta}_{ik} - \tilde{\theta}_{i0} - \tilde{\theta}_{0k},
\end{eqnarray}
where corresponding dual affine coordinate becomes $\eta_{i}=\sum_{k'=0}^{m} g_{ik'}$, $\eta'_{k}=\sum_{i'=0}^{n} g_{i'k}$ and $\eta_{ik}=g_{ik}$. With these definitions, hereinafter we employ the mixed coordinate of $\zeta = \left( \Xi, \Theta \right)$ where $\Xi=\left\{ \eta_{1},\cdots,\eta_{n},\eta'_{1},\cdots,\eta'_{m} \right\}$ and $\Theta=\left\{ \theta_{00},\cdots, \theta_{nm} \right\}$ for convenience.
Then we introduce separable canonical distribution derived from configuration $\mathbf{Q}$ as
\begin{eqnarray}
g_{\textrm{c0}}^{\mathbf{Q}} \left( x,y \right) = \sum_{y} g_{\textrm{c}}^{\mathbf{Q}} \left( x,y \right) \cdot \sum_{x} g_{\textrm{c}}^{\mathbf{Q}} \left( x,y \right).
\end{eqnarray}
Under these preparations, it is now clear from dual orthogonality\cite{ig} that $m$-geodesic between $g_{\textrm{c}}^{\mathbf{Q}}$ and $g_{\textrm{c0}}^{\mathbf{Q}}$ is orthogonal to $e$-geodesic between $g_{\textrm{c0}}^{\mathbf{Q}}$ and $g_{\textrm{G}}^{\mathbf{Q}}$, since $m$-flat subspace $M$ with linear constraints for $\Xi$ of
\begin{eqnarray}
\label{eq:af0}
\forall i,k:\quad \eta_{i} &=& \sum_{k}g_{\textrm{c}}\left( x_{i},y_{k} \right), \eta'_{k}=\sum_{i}g_{\textrm{c}}\left( x_{i},y_{k} \right)
\end{eqnarray}
contains the elements of $g_{\textrm{c}}^{\mathbf{Q}}$ and $g_{\textrm{c0}}^{\mathbf{Q}}$ (i.e., the same marginal distribution as $g_{\textrm{c}}^{\mathbf{Q}}$), and $e$-flat subspace $I$ with linear constraints for $\Theta$ of
\begin{eqnarray}
\label{eq:af1}
\forall i,k:\quad \theta_{ik} &=& 0
\end{eqnarray}
contains the elements of $g_{\textrm{c0}}^{\mathbf{Q}}$ and $g_{\textrm{G}}^{\mathbf{Q}}$ (i.e., separable distributions).
Therefore, by employing generalized Pythagorean theorem,\cite{ig} we first get
\begin{eqnarray}
\label{eq:KL}
D_{\textrm{KL}}\left(  g_{\textrm{c}}^{\mathbf{Q}} : g_{\textrm{G}}^{\mathbf{Q}} \right) &=& {D_{\textrm{KL}}} \left( g_{\textrm{c}}^{\mathbf{Q}} : g_{\textrm{c0}}^{\mathbf{Q}} \right) + {D_{\textrm{KL}}} \left( g_{\textrm{c0}}^{\mathbf{Q}} : g_{\textrm{G}}^{\mathbf{Q}} \right), 
\end{eqnarray}
which indicates that local nonlinearity in $\phi$ derived from configuration $\mathbf{Q}$ (l.h.s.) can be straightforwardly decomposed in terms of the sum of Kullback-Leibler (KL) Divergence\cite{KL} of non-separability for canonical distribution (hereinafter call canonical non-separability of the 1st term in r.h.s.) and deviation from Gaussian-derived distribution (the 2nd term of r.h.s.). Here, by introducing statistical manifold, SDF-based local information of $D_{\textrm{KL}}\left( g_{\textrm{c}}^{\mathbf{Q}} : g_{\textrm{c0}}^{\mathbf{Q}} \right)$ can be naturally extracted from local nonlinearity. We emphasize again that this decomposition is essentially independent of temperature and energy, and depends only on CDOS landscape and given configuration, which has the same desired characteristics as ASDF. 
However, Eq.~\eqref{eq:KL} provides insufficient information about local nonlinearity in terms of ASDF: For instance, when we focus on decomposition at origin of configuration space (corresponding to perfectly disordered state), ASDF clearly takes zero vector, which means that local nonlinearity becomes zero at the disordered state in terms of ASDF. Meanwhile, at the origin of $\mathbf{Q}_{0}$, since $g_{\textrm{c}}^{\mathbf{Q}_{0}}$ and $g_{\textrm{G}}^{\mathbf{Q}_{0}}$ are respectively identical to practical CDOS and Gaussian themselves, $D_{\textrm{KL}}\left(  g_{\textrm{c}}^{\mathbf{Q}_{0}} : g_{\textrm{G}}^{\mathbf{Q}_{0}} \right)\neq 0$, which exhibits different behavior from ASDF. 
This is naturally accepted since ASDF focuses on difference in center of gravity (COG) between $g_{\textrm{c}}^{\mathbf{Q}_{0}}$ and 
$g_{\textrm{G}}^{\mathbf{Q}_{0}}$: Therefore, $D_{\textrm{KL}}\left(  g_{\textrm{c}}^{\mathbf{Q}_{0}} : g_{\textrm{G}}^{\mathbf{Q}_{0}} \right)$ naturally contains information about local nonlinearity in $\phi$, which is not included in local nonlinearity information obtained from ASDF. 

From above discussions, we require additional measure for local nonlinearity in order to clarify how previously-proposed ASDF-based nonlinearity is interpreted in terms of the present expression of nonlinearity on statistical manifold. For this purpose, we introduce an additional  
distribution of
\begin{eqnarray}
\label{eq:origin}
g_{\textrm{c}}^{'\mathbf{Q}} \left( x,y \right)= g_{\textrm{c}}^{\mathbf{Q}} \left( x - \sum_{x,y}x\cdot g_{\textrm{G}}^{\mathbf{Q}}\left( x,y \right), y - \sum_{x,y}y\cdot g_{\textrm{G}}^{\mathbf{Q}}\left( x,y \right)    \right).
\end{eqnarray}
This distribution of $g_{\textrm{c}}^{'\mathbf{Q}}$ is selected so that it satisfies two important characteristics of (i) $D_{\textrm{KL}}\left(  g_{\textrm{c}}^{'\mathbf{Q}} : g_{\textrm{G}}^{\mathbf{Q}} \right)$ corresponds to \textit{ASDF-independent} local nonlinearity information derived from $\mathbf{Q}$, and (ii) at the origin of $\mathbf{Q}_{0}$, $g_{\textrm{c}}^{\mathbf{Q}_{0}} = g_{\textrm{c}}^{'\mathbf{Q}_{0}}$. 
\begin{figure}[h]
\begin{center}
\includegraphics[width=0.72\linewidth]{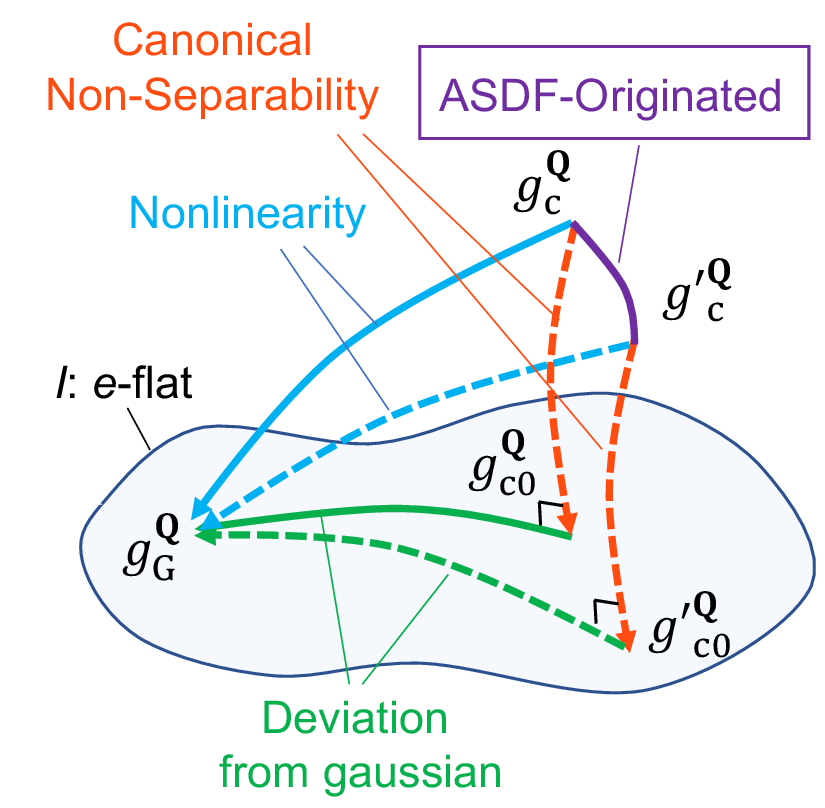}
\caption{ Geometric interpretation of local nonlinearity in $\phi$ derived from configuration $\mathbf{Q}$ for binary alloys with pair correlations. Solid and dashed curves respectively correspond to ASDF-dependent and -independent contribution.  }
\label{fig:dec}
\end{center}
\end{figure}
In a similar fashion to Eq.~\eqref{eq:KL}, we can also apply generalized Pythagorean theorem for difference between $g_{\textrm{c}}^{'\mathbf{Q}}$ and $g_{\textrm{G}}^{\mathbf{Q}}$, namely
\begin{eqnarray}
\label{eq:KL2}
D_{\textrm{KL}}\left(  g_{\textrm{c}}^{'\mathbf{Q}} : g_{\textrm{G}}^{\mathbf{Q}} \right) &=& {D_{\textrm{KL}}} \left( g_{\textrm{c}}^{'\mathbf{Q}} : g_{\textrm{c0}}^{'\mathbf{Q}} \right) + {D_{\textrm{KL}}} \left( g_{\textrm{c0}}^{'\mathbf{Q}} : g_{\textrm{G}}^{\mathbf{Q}} \right), \nonumber \\
\quad 
\end{eqnarray}
where 
$g_{\textrm{c0}}^{'\mathbf{Q}} \left( x,y \right) = \sum_{y} g_{\textrm{c}}^{'\mathbf{Q}} \left( x,y \right) \cdot \sum_{x} g_{\textrm{c}}^{'\mathbf{Q}} \left( x,y \right)$.
Hereinafter, the primes on distribution always denotes that their center of gravity takes the same as corresponding Gaussian-derived canonical distribution. From above discussions, it is now clear that decomposition of local nonlinearity in Eq.~\eqref{eq:KL} is ASDF-dependent, while that in Eq.~\eqref{eq:KL2} is ASDF-independent: Geometric interpretations for local nonlinearity in $\phi$ derived from $\mathbf{Q}$ is summarized in Fig.~\ref{fig:dec}. From Fig.~\ref{fig:dec} and the character of KL divergence, we see that canonical non-separability is clearly independent of ASDF in this decomposition: ASDF-originated contribution to the local nonlinearity is restricted to contribution from differences in separable distribution from Gaussian. This fact and the character of global nonlinearity in tropical limit indicate that information about \textit{evolution} of ASDF (as dynamical system) should be required to include contribution from canonical non-separability.
Note that ASDF-dependent local nonlinearity of $D_{\textrm{KL}}\left(  g_{\textrm{c}}^{\mathbf{Q}} : g_{\textrm{G}}^{\mathbf{Q}} \right)$ is not purely originated from ASDF, and one of the natural choice for counterpart of the ASDF on the manifold can be minimal path between $g_{\textrm{c}}^{\mathbf{Q}}$ and $g_{\textrm{c}}^{'\mathbf{Q}}$ since ASDF connects two configurations at the minimal path: The corresponding length on the manifold is given by $2\cos^{-1}\left( \sqrt{g_{\textrm{c}}^{\mathbf{Q}}}\cdot \sqrt{g_{\textrm{c}}^{'\mathbf{Q}}} \right)$,\cite{path} where the dot denotes inner product and $\sqrt{g}:=\left( \sqrt{g_{00}}, \cdots \sqrt{g_{nm}} \right)$.

\begin{figure}[h]
\begin{center}
\includegraphics[width=0.83\linewidth]{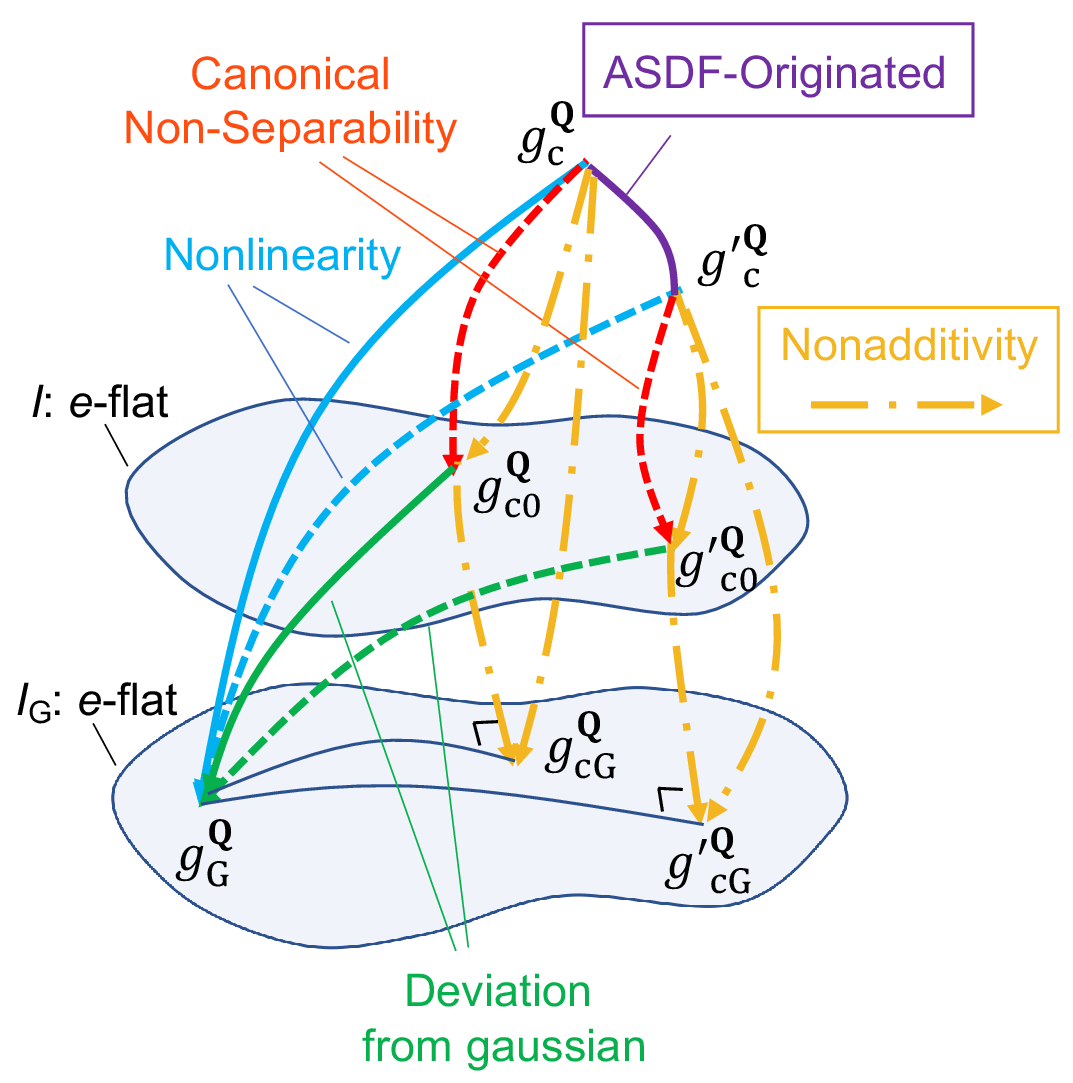}
\caption{Generalized geometric interpretation for local nonlinearity in $\phi$ for multicomponent alloys. Solid and dashed curves respectively correspond to ASDF-dependent and -independent contribution. }
\label{fig:dec2}
\end{center}
\end{figure}
Next, we extend the above discussions to any system where $\Gamma$ is not diagonal: This holds for binary $\left(R=2\right)$ systems with multisite (including 3- and higher-body) correlations and any multicomponent $\left(R\ge 3\right)$ systems. In these cases, canonical distribution for linear system derived from Gaussian CDOS ($g_{\textrm{G}}^{\mathbf{Q}}$) is no longer separable, i.e., 
$g_{\textrm{G}}^{\mathbf{Q}} \notin I,$
which means that we cannot straightforwardly apply generalized Pythagorean theorem to decompose nonlinearity as shown in Fig.~\ref{fig:dec}. 
In order to address this problem, we apply dual-orthogonal foliation\cite{ig} to statistical manifold $S$ with a set of linear constraint of $\Theta = \mathbf{F}$ ($\mathbf{F}$ denotes $\left( n\cdot m +1 \right)$-component real vector), resulting in a family $\mathbf{I}_{F}$ consisting of $e$-flat subspaces with $I\in\mathbf{I}_{F}$. 
Therefore, we can find the linear constraint:
\begin{eqnarray}
\exists \mathbf{F}_{\textrm{G}}\quad \mathrm{s.t.}\quad g_{\textrm{G}}^{\mathbf{Q}} \in I_{{\textrm{G}}} \in \mathbf{I}_{F}
\end{eqnarray}
from Eq.~\eqref{eq:af}. Meanwhile, when we consider $m$-flat subspace $M$ with linear constraint for $\Xi$ of Eq.~\eqref{eq:af0}, $M$ is orthogonal to $I_{{\textrm{G}}}$ from dual orthogonality, where $M$ and $I_{{\textrm{G}}}$ have one common distribution $g_{\textrm{CG}}^{\mathbf{Q}}$. 
This fact immediately leads to decomposition of nonlinearity as 
\begin{eqnarray}
\label{eq:KL3}
{D_{\textrm{KL}}}\left(  g_{\textrm{C}}^{\mathbf{Q}} : g_{\textrm{G}}^{\mathbf{Q}} \right) &=& {D_{\textrm{KL}}} \left( g_{\textrm{C}}^{\mathbf{Q}} : g_{\textrm{CG}}^{\mathbf{Q}} \right) + {D_{\textrm{KL}}} \left( g_{\textrm{CG}}^{\mathbf{Q}} : g_{\textrm{G}}^{\mathbf{Q}} \right),  \nonumber \\
\quad
\end{eqnarray}
which provides implicit information about nonlinearity in terms of constraints to SDFs. 
From above discussions, since $e$-geodesic between $g_{\textrm{G}}^{\mathbf{Q}}$ and $g_{\textrm{CG}}^{\mathbf{Q}}$ is orthogonal to $m$-geodesic between $g_{\textrm{CG}}^{\mathbf{Q}}$ and $g_{\textrm{C0}}^{\mathbf{Q}}$ at $g_{\textrm{CG}}^{\mathbf{Q}}$, we further apply generalized Pythagorean theorem to rewrite $D_{\textrm{KL}}\left( g_{\textrm{G}} : g_{\textrm{CG}} \right)$ in terms of the canonical non-separability in a similar fashion to Fig.~\ref{fig:dec}. 
Then we can obtain generalized geometric decomposition of nonlinearity in $\phi$  as follows:
\begin{eqnarray}
\label{eq:klf}
{D_{\textrm{KL}}}\left(  g_{\textrm{C}}^{\mathbf{Q}} : g_{\textrm{G}}^{\mathbf{Q}} \right) = {D_{\textrm{KL}}} \left( g_{\textrm{C}}^{\mathbf{Q}} : g_{\textrm{C0}}^{\mathbf{Q}} \right) + {D_{\textrm{KL}}}\left( g_{\textrm{C0}}^{\mathbf{Q}} : g_{\textrm{G}}^{\mathbf{Q}} \right) + \Delta D_{\textrm{NAE}}^{\mathbf{Q}}, \nonumber \\
\quad
\end{eqnarray}
where 
\begin{eqnarray}
\label{eq:nae}
\Delta D_{\textrm{NAE}}^{\mathbf{Q}}  = {D_{\textrm{KL}}}\left(  g_{\textrm{C}}^{\mathbf{Q}}  : g_{\textrm{CG}}^{\mathbf{Q}}  \right) -  {D_{\textrm{KL}}}\left(  g_{\textrm{C}}^{\mathbf{Q}}  : g_{\textrm{C0}}^{\mathbf{Q}}  \right) - {D_{\textrm{KL}}}\left(  g_{\textrm{C0}}^{\mathbf{Q}}  : g_{\textrm{CG}}^{\mathbf{Q}}  \right)  . \nonumber \\
\quad
\end{eqnarray}
The first term of r.h.s. of Eq.~\eqref{eq:klf} corresponds to canonical non-separability in SDF, 
the second term to contribution from deviation from Gaussian in analogy to Fig.~\ref{fig:dec}, and the third term 
$\Delta D_{\textrm{NAE}}$ represents nonadditive character for non-separability in SDF as clearly seen from Eq.~\eqref{eq:nae}.
For systems with diagonal $\Gamma$, $\Delta D_{\textrm{NAE}}$ can take zero for any configuration, while for those with non-diagonal $\Gamma$,  $\Delta D_{\textrm{NAE}}$ cannot generally be zero (but it can take zero for specific set of configurations, where $m$-geodesic between $g_{\textrm{C}}$ and $g_{\textrm{C0}}$, and $e$-geodesic between 
$g_{\textrm{C0}}$ and $g_{\textrm{CG}}$, are orthogonal at $g_{\textrm{C0}}$, and any deviation from this condition leads to positive or negative contribution to the nonlinearity).
In order to provide clear interpretation for the generalized decomposition based on ASDF, we can employ another type of decomposition 
by changing origin for distributions of $g_{\textrm{C}}^{\mathbf{Q}}$, $g_{\textrm{C0}}^{\mathbf{Q}}$ and $g_{\textrm{CG}}^{\mathbf{Q}}$ 
according to the same coordinate transform of Eq.~\eqref{eq:origin}, respectively resulting in $g_{\textrm{C}}^{'\mathbf{Q}}$, $g_{\textrm{C0}}^{'\mathbf{Q}}$ and $g_{\textrm{CG}}^{'\mathbf{Q}}$.  Then, in a similar fashion to the decomposition in Fig.~\ref{fig:dec}, we obtain generalized geometric interpretation for local nonlinearity in $\phi$ in terms of ASDF, given by Fig.~\ref{fig:dec2}.
From Fig.~\ref{fig:dec2}, it is now clear that in addition to the canonical non-separability, the nonadditivity is also an ASDF-independent contribution to local nonlinearity in $\phi$.
Since (i) canonical non-separability in SDF is specific to $R\ge 2$ systems under pair correlations, and (ii) nonadditivity in the non-separability is specific to other systems, analyzation of the non-separability as well as its nonadditivity, 
$\Delta D_{\textrm{NAE}}$, should play key roles for understanding nonlinearity in $\phi$ for multicomponent alloys. 
We finally note that while the present study provide geometric intuition of local nonlinearity regarding previously-introduced vector field of ASDF, it is still unclear how the nonlinearity as ASDF and nonlinearity as divergence on statistical manifold are connected through concrete formulations, since they are independently defined on different space: Our recent study reveal that the local nonlinearity on two different worlds can be bridged through stochastic thermodynamics with introducing special transition,\cite{st} which should be further examined in our future study.


\section{Conclusions}
By introducing statistical manifold, we here address how local nonlinearity in canonical average is interpreted by previously-introduced vector field $\mathbf{A}$ and constraints to structural degree of freedoms. The results show that the local nonlinearity can be decomposed into sum of the two (for binary alloys with pair correlations) or three (for otherwise) contributions in terms of the Kullback-Leibler divergence, where this decomposition is independent of temperature and many-body interaction. $\mathbf{A}$-dependent and $\mathbf{A}$-independent decomposition of local nonlinearity indicates that information about \textit{evolution} of the vector field should be required to address the non-separability and nonadditivity. The present results provide how configuration-dependent local nonlinearity for multicomponent alloys is dominated in terms of constraints SDFs, which will play significant role to design nonlinearity through changing underlying lattice geometry and compositions.

\section{Acknowledgement}
This work was supported by Grant-in-Aids for Scientific Research on Innovative Areas on High Entropy Alloys through the grant number JP18H05453 and a Grant-in-Aid for Scientific Research (16K06704) from the MEXT of Japan and Research Grant from Hitachi Metals$\cdot$Materials Science Foundation.


\begin{thebibliography}{9}
\bibitem{ce} J.M. Sanchez, F. Ducastelle, and D. Gratias, Physica A \textbf{128}, 334 (1984).
\bibitem{mc1} N. Metropolis, A. W. Rosenbluth, M. N. Rosenbluth, A. H. Tellerand, and E. Teller, J. Chem. Phys. \textbf{21}, 1087 (1953).
\bibitem{mc2} J. Lee, Phys. Rev. Lett. 71, 211 (1993).
\bibitem{wl} F. Wang and D.P. Landau, Phys. Rev. Lett. \textbf{86}, 2050 (2001).
\bibitem{bd} K. Yuge and S. Ohta, J. Phys. Soc. Jpn. \textbf{88}, 104803 (2019).
\bibitem{trop} K. Yuge and S. Ohta, J. Phys. Soc. Jpn. \textbf{89}, 084802 (2020).
\bibitem{em2} K. Yuge, J. Phys. Soc. Jpn. \textbf{85}, 024802  (2016).
\bibitem{asdf} K. Yuge, J. Phys. Soc. Jpn. \textbf{86}, 104802 (2018).
\bibitem{ig} S. Amari, \textit{Information Geometry and Its Applications}, Springer (2016).
\bibitem{mom} S. Ohta and K. Yuge, J. Phys. Soc. Jpn. \textbf{88}, 034802  (2019).
\bibitem{KL} T. M. Cover and J. A. Thomas, \textit{Elements of information theory} (John Wiley \& Sons, 2012).
\bibitem{path} W. K. Wootters, Phys. Rev. D, \textbf{23}, 357 (1981).
\bibitem{st} K. Yuge,  	arXiv:2103.12414 [cond-mat.stat-mech].
\end{thebibliography}
\end{document}